\pdfoutput=1
\documentclass{scrartcl}

\usepackage{amssymb}
\usepackage{comment}
\usepackage{amsmath}
\usepackage{bookmark}
\usepackage{hyperref}
\usepackage{tensor}
\usepackage{graphicx}
\usepackage{caption}
\usepackage{subcaption}
\usepackage[title]{appendix}
\newcommand{\eqnref}[1]{Eq.~(\ref{#1})}
\newcommand{\LUF}{\mathrm{L}_{\mathcal{F}} \mathrm{U}(1)} 

\title{'t Hooft anomalies in metals}
\author{Dominic V. Else \\ Perimeter Institute for Theoretical Physics}
\date{February 24, 2025}

\usepackage[style=phys,biblabel=brackets,eprint]{biblatex}
\begin{document}
\maketitle
\begin{abstract}
I review some recent results on understanding the physics of metals in an exact non-perturbative way through the powerful field-theoretic concepts of emergent symmetries and 't Hooft anomalies. A 't Hooft anomaly is a discrete topological property that quantum field theories with global symmetries can have. I explain how many of the properties of metals can in fact be viewed as direct consequences of the anomaly. This allows a structural understanding of metals, including non-Fermi liquids, to be obtained even in the absence of any exact solution for the strongly coupled dynamics. I then outline the main limitations and outstanding questions.
\end{abstract}
\section{Introduction}
Understanding the physics of strongly coupled quantum many-body systems is a challenging task in theoretical physics. Nowhere is this challenge more acute than in the study of so-called ``non-Fermi liquid metals''. A metal is, roughly, a state of matter in which electrons can flow freely even at low temperatures. The simplest example of a metal is a non-interacting Fermi gas, in which the electrons occupy single-particle states that are labeled by their momentum, giving rise to a Fermi surface. The effect of interactions can be taken into account through Fermi liquid theory \cite{Landau__1957,FermiLiquidBook,Polchinski_9210,Shankar_9307}, in which the bare electrons near the Fermi surface become dressed to form long-lived quasiparticles, in terms of which the low-energy physics can be described.
By contrast, in non-Fermi liquid metals, the quasiparticles are destroyed and the system forms a more strongly interacting quantum many-body state.
The fact that not all metals are well characterized by Fermi liquid theory has become apparent experimentally \cite{Proust_1807,Varma_1908,Lohneysen_0606,Gegenwart_0712}.

Strongly coupled systems such as non-Fermi liquids pose a challenge for traditional modes of doing theoretical condensed matter physics. For example, doing perturbation theory in the interaction strength will be futile. After intense theoretical effort, models of fermions coupled to a critical boson have been found in which results can be obtained perturbatively in some (artificial) parameter such as the inverse of the number of fermion flavors, allowing some non-Fermi liquid physics to be obtained (e.g.\ see Ref.~\cite{Lee_1703}), though one must still ask whether such models can capture the physics seen in experiment.

In this review, I will describe some recent results coming at the question of how to understand the physics of metals, including non-Fermi liquid metals, from a different angle. Specifically, the question is whether it is possible to describe at least some part of the behavior of general metals in a fully exact and non-perturbative way. To do this, we can exploit general results in quantum field theory, specifically the concept of a `` 't Hooft anomaly'' \cite{tHooftAnomaly, Kapustin_1403}. A 't Hooft anomaly is a discrete, exact, topological non-perturbative property of a quantum field theory with a global symmetry. Roughly, it can be viewed as an obstruction to gauging the global symmetry. In the past 15 years or so, the general classification of 't Hooft anomalies has been better understood by exploiting a bulk-boundary correspondence with the classification of symmetry-protected topological (SPT) phases of matter \cite{Senthil_1405}.

One way to view Fermi liquid theory is that it corresponds to an ``effective field theory'' which describes the low-energy physics of a metal \cite{Polchinski_9210,Shankar_9307}. One of the main insights I will describe in this review is that this effective field theory in fact has a 't Hooft anomaly for an emergent symmetry, and many (though not all) of the properties of Fermi liquid theory can be understood as consequences of this anomaly. However, the main strength of the anomaly perspective is that it has the potential to apply equally well to non-Fermi liquid metals, allowing certain properties of non-Fermi liquid metals to be described in an exact non-perturbative way.


The rest of this review is structured as follows. In Section \ref{sec:introduction_anomalies}, I give a brief introduction to 't Hooft anomalies using some simple examples. In Section \ref{sec:loop_group}, I describe the emergent symmetry group of metals. In Section \ref{sec:loop_group_anomalies}, I describe the structure of the 't Hooft anomalies of this emergent symmetry group. In Section \ref{sec:beyond_fermi}, I discuss the applicability of these ideas to non-Fermi liquids. In Section \ref{sec:consequences}, I show how many of the properties of metals can be understood in terms of the anomaly. In Section \ref{sec:limitations}, I describe some limitations and extensions of these results, as well as the possible tension with experimentally observed $T$-linear resistivity (Section \ref{subsec:conundrum}). In Section \ref{sec:beyond_metals}, I mention some analogous results for non-metallic systems. Finally, in Section \ref{sec:outlook}, I conclude and discuss future directions.

\section{Introduction to 't Hooft anomalies}
\label{sec:introduction_anomalies}

\subsection{Helical fermions in one spatial dimension}
\label{eq:helical_fermions}
A 't Hooft anomaly is a property that a quantum field theory with global symmetry $G$ can have. As a simple example of a 't Hooft anomaly, let us consider a free-fermion system in one dimension with conservation of both charge and the $z$ component of spin (for brevity we will henceforth refer to the $z$ component of spin as just ``spin''). Thus, we can write the symmetry group as $G = \mathrm{U}(1)_\uparrow \times \mathrm{U}(1)_\downarrow$, where the generators of $\mathrm{U}(1)_{\uparrow/\downarrow}$ are the total number $N_{\uparrow/\downarrow}$ of spin-up/spin-down particles respectively. The total fermion number is $Q = N_\uparrow + N_\downarrow$.

\begin{figure}
\includegraphics[scale=0.7]{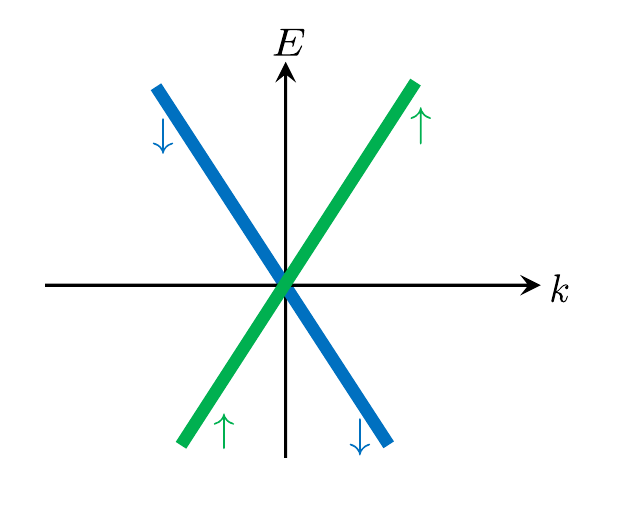}
\caption{\label{fig:helical}The dispersion relation for a non-interacting helical fermion in 1D. The spin-up particles move to the right while the spin-down particles move to the left.}
\end{figure}
Let us suppose that the low-energy physics is described by two non-interacting fermion bands with opposite chirality, where each band is spin-polarized, as shown in Figure \ref{fig:helical}: the right-moving band carries only spin-up, while the left-moving band carries only spin down. I claim that this system will carry a 't Hooft anomaly for the $\mathrm{U}(1)_\uparrow \times \mathrm{U}(1)_\downarrow$ symmetry. The most intuitive manifestation of this is that it is impossible to give the system a spatial boundary without breaking the symmetry. Indeed, if we impose charge conservation at the boundary, then a right-moving particle with spin-up that is incident on the boundary must get reflected back to a left-moving particle with spin-down, which breaks spin conservation.

It is important to note that the anomaly is a property that is robust to interactions, even very strong interactions that drive the system into a completely different phase, as long as they respect the global symmetry. Indeed, suppose that by adding symmetry-respecting interactions it were possible to drive the system into a phase which does support a boundary that respects the symmetry. Then we could simply switch on the interaction at position $x > 0$ and then have a boundary at some $x=L$, effectively constructing a boundary for the free-fermion system at $x < 0$, which is disallowed by the anomaly.

One way to think about this result is that the anomaly is not really a statement about the Hamiltonian of the quantum field theory. Rather, it is simply a statement about how the symmetry acts on the Hilbert space. In particular, what one can show in the present example is that if we write the generators of the $\mathrm{U}(1)_\uparrow \times \mathrm{U}(1)_\downarrow$ symmetry as integrals of local densities, i.e.\
\begin{equation}
    N_\sigma = \int n_\sigma(x) dx,
    \end{equation}
    (where $\sigma$ can be $\uparrow$ or $\downarrow$),
    then the anomaly corresponds to the statement that the local densities fail to commute, specifically:
    \begin{align}
        \label{eq:densities_commutator}
    [n_\sigma(x), n_\sigma(x')] &= \frac{i m_\sigma}{2\pi} \delta'(x-x'), \\
    [n_\uparrow(x), n_\downarrow(x')] &= 0.
    \end{align}
    where $\delta'$ is the derivative of the Dirac delta function, and $m_\uparrow = 1$, $m_\downarrow = -1$.

A powerful way to probe the anomaly that can readily be generalized to higher dimensions is to apply a background gauge field for the global symmetry. In the case under consideration, we can just apply an electric field [which formally is a background gauge field for the charge $\mathrm{U}(1)$, i.e.\ the diagonal subgroup of $\mathrm{U}(1)_\uparrow \times \mathrm{U}(1)_\downarrow$]. What one will find is that due to the 't Hooft anomaly, the conservation law for spin becomes violated in the presence of electric field. Specifically, let us define a space-time vector $j^{\mu \, \sigma}$ whose time and spatial component are the charge density and current respectively for the $N_\sigma$ charge ($\sigma = \uparrow/\downarrow$). [In the rest of the paper, for brevity we will always refer to the space-time vector $j^\mu$ asociated to a conserved quantity as the ``current''; thus, its time component is the density of the conserved quantity and its spatial components are the spatial current.] Then, in response to an electric field $E$ one obtains the modified conservation law
\begin{equation}
    \label{eq:anomaly_1d}
\partial_\mu j^{\mu \, \sigma} = \frac{m_\sigma}{2\pi} E
\end{equation}
    The fact that the right-hand side is nonzero shows that $N_\uparrow$ and $N_\downarrow$ are not separately conserved in the presence of an electric field. Note that anomaly equations such as \eqnref{eq:anomaly_1d} are generally very rigid: one can show from the quantization of charge and spin that the coefficients $m_\sigma$ have to be integers, and it is not possible to deform the right-hand side of \eqnref{eq:anomaly_1d} by adding additional terms. \eqnref{eq:anomaly_1d}, in the sense that it is invariant under space-time diffeomorphisms and does not depend on the metric.

In general, the anomalies for a given global symmetry in a given spatial dimension are classified by a discrete Abelian group. The group operation can be interpreted as ``stacking'': for example, if we considered \emph{two} copies of our free-fermion system, then we would have $m_{\uparrow/\downarrow} = +2/-2$ in \eqnref{eq:anomaly_1d}. For the case of $\mathrm{U}(1)_\uparrow \times \mathrm{U}(1)_\downarrow$ symmetry, the anomalies are classified by an integer $m = m_\uparrow = -m_\downarrow$, so the classification group is the additive group of integers, ``$\mathbb{Z}$''.

Another consequence of the anomaly is that it enforces the gaplessness of the ground state. More precisely, the statement we can make for the current example is that the anomaly \emph{necessarily} requires both charge and spin to be gapless. To see this, note if there  were a non-zero spin gap, then a weak electric field would be unable to overcome the gap to create excitations carrying spin, and so it would be impossible for \eqnref{eq:anomaly_1d} to be satisfied. Meanwhile, if there were a non-zero charge gap, then at low energies the system should not feel the electric field at all, and again it would be impossible for \eqnref{eq:anomaly_1d} to be satisfied. To illustrate the enforced gaplessness, observe that for example, free-fermion scattering terms between the left-moving and right-moving band, which would open up a gap if they were allowed, are disallowed by conservation of spin. The fact that \emph{no} term, even with interactions, can be added to the Hamiltonian to gap out the system without breaking the symmetry is a non-trivial consequence of the anomaly.

Another interesting point about 't Hooft anomalies is that they are in one-to-one correspondence with the classification of symmetry-protected topological (SPT) phases in one higher dimension. For example, the anomaly described in this section appears on the boundary of the so-called ``quantum spin Hall'' states in 2D \cite{Kane_0411}. These phases are characterized by the property that applying an electric field induces a spin current in the direction perpendicular to the electric field. The total spin of the 2D system, including its boundary, is conserved; one views \eqnref{eq:anomaly_1d} as reflecting the inflow of spin from the 
2D bulk.

A final consequence of 't Hooft anomalies is that they represent an obstruction to realizing the field theory as the low-energy limit of a $d$-dimensional lattice system with the symmetry $G$ implemented microscopically and commuting with the Hamiltonian on the lattice\footnote{Note that while the impossibility to realize the system on the lattice will be true for the 't Hooft anomalies considered in this paper (see, for example, Ref.~\cite{Kapustin_2401}), the strongest version of this statement is not true for other 't Hooft anomalies. For example, in many cases when the symmetry group $G$ is finite, the system can be realized on the lattice, albeit at the cost of introducing so-called ``non-on-site'' microscopic symmetries in which the symmetry generators cannot be written as a tensor-product over individual lattice site \cite{Chen_1106, Else_1409}.}.
To see this intuitively, observe that, for example, it would be impossible to extend the bands shown in Figure \ref{fig:helical} to a periodic Brillouin zone while maintaining spin as a good quantum number for the bands. To argue that this statement holds even with interactions one can note that, for example, the equation \eqnref{eq:densities_commutator} cannot be defined in a consistent way on the lattice. Of course, the discussion in the previous paragraph shows that the field theory \emph{can} represent the boundary theory of a $d+1$-dimensional lattice system.

    \subsection{'t Hooft anomalies of emergent symmetries: one-dimensional metals}
    An important point is that in discussing 't Hooft anomalies, we do not need to confine ourselves to the symmetries of the microscopic Hamiltonian (e.g. the Hamiltonian of electrons hopping on the lattice). We can instead consider the symmetries of the effective field theory which describes the low-energy physics of the system. This can include symmetries which do not correspond to any symmetry of the microscopic Hamiltonian; these are referred to as ``emergent symmetries''.

    Recall from the previous section that a 't Hooft anomaly of a symmetry represents an obstruction to realizing the symmetry microscopically on the lattice. However, for emergent symmetries, which are not \emph{supposed} to be implemented microscopically, this is perfectly acceptable. Thus, the quantum field theory under consideration can emerge as the low-energy effective theory of a microscopic Hamiltonian on the lattice (without needing to be on the boundary of a bulk system in one higher dimension) even when it has a 't Hooft anomaly for an emergent symmetry.

    \begin{figure}
    \includegraphics[scale=0.7]{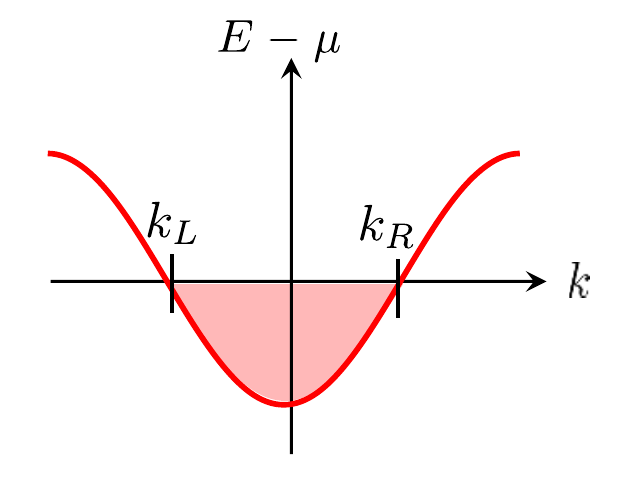}
    \caption{\label{fig:1d_metal}The dispersion relation for a non-interacting 1-D metal.}
    \end{figure}

 As an example of this phenomenon that is particularly relevant for this review article, we will consider the case of a one-dimensional metal, which will serve as a good warm-up for the higher-dimensional metals that we will discuss later. For simplicity, one could consider the case of non-interacting metals, though the same considerations will apply to interacting Luttinger liquids as well. In a one-dimensional metal, the low-energy physics separately conserves ``left-moving'' and ``right-moving'' charge, i.e. the charge near $k_L$ and $k_R$ in Figure \ref{fig:1d_metal}. We will denote these by $N_L$ and $N_R$. We note that apart from the fact that left-moving and right-moving branches are at different momenta, the low-energy physics is now identical to the theory considered in Section \ref{eq:helical_fermions}, with $N_L$ and $N_R$ replacing $N_\uparrow$ and $N_\downarrow$. In particular, this means that the emergent $\mathrm{U}(1)_L \times \mathrm{U}(1)_R$ will have a 't Hooft anomaly, with the same structure that we already considered in Section \ref{eq:helical_fermions}. This is the 1D analog of the anomalies of metals in higher dimensions that we will shortly consider.

\subsection{Anomalies and bulk topological terms}
We have already mentioned the fact that anomalies of a symmetry $G$ in $d-1$ spatial dimensions are in correspondence with SPT phases protected by symmetry $G$ in $d$ spatial dimensions. A very general way to characterize SPT phases, and hence anomalies, is through the formalism of ``topological terms''. Imagine taking a system in an SPT phase in $d$ spatial dimensions and coupling it to a background gauge field $A$ for the $G$ symmetry. Since SPT phases are gapped, we can integrate out all the dynamical fields. Then the only thing that can be left\footnote{Technically this requires something more than just that the system be gapped; the system must be \emph{trivially} gapped, in the sense that there are no topological dynamical degrees of freedom remaining after integrating out the gapped degrees of freedom. This is true for SPT phases but not for so-called ``non-invertible'' or ``intrinsic'' topological phases such as the toric code.} in the action is a term involving the background gauge field $A$. In general this will be topological (i.e.\ it is invariant under diffeomorphisms of space-time and does not depend on the space-time metric.) It should also be invariant under gauge transformations of the gauge field. By classifying such topological terms, one can classify SPT phases.

For example, consider the case $d=2$, $G = \mathrm{U}(1)$. Then the SPT phases are the integer quantum Hall states\footnote{Technically these are not SPT phases because they actually still a have a non-trivial ``gravitational'' response even when the $\mathrm{U}(1)$ symmetry is broken. We will disregard this subtlety here.}. The corresponding topological term is the so-called ``Chern-Simons term'':
\begin{equation}
    \label{eq:chern_simons}
    \frac{m}{4\pi} \int \epsilon^{\mu \nu \lambda} A_\mu \partial_\nu A_\lambda d^3 x,
\end{equation}
where $A$ is a $\mathrm{U}(1)$ gauge field. One can show that such a term is gauge-invariant as long as the coefficient $m$ is quantized to be an integer \cite{Witten_1510}. One can convince oneself that such terms indeed encode the appropriate physics for integer quantum Hall states; for example, by computing the spatial current $j^i = \frac{\delta S}{\delta A_i}$ [where we use Roman letters $i,j$, etc. to denote spatial indices], we find
\begin{equation}
    \label{eq:Hall_current}
    j^i = \frac{m}{2\pi} \epsilon^{ij} E_j,
\end{equation}
where $E_j = \partial_t A_i - \partial_j A_t$ is the electric field. This shows that the coefficient $m$ corresponds to the Hall conductance. Moreover, in general if the system has a boundary, then \eqnref{eq:Hall_current} shows that in the presence of an electric field in the direction parallel to the boundary, there will be charge flowing onto the boundary. Therefore, the charge of the boundary will be non-conserved, leading to the modified conservation law for the 1D boundary theory:
\begin{equation}
    \partial_\mu j^\mu = \frac{m}{2\pi} E,
\end{equation}
[compare \eqnref{eq:anomaly_1d}], which is a reflection of the anomaly of the boundary theory.

To avoid cumbersome notation, here and in the rest of this review we prefer to write topological terms such as \eqnref{eq:chern_simons} in a coordinate-free representation in the language of differential forms \cite{Nakahara}, in which case \eqnref{eq:chern_simons} becomes
\begin{equation}
    \frac{m}{4\pi} \int A \wedge dA.
\end{equation}

\section{``Loop group'' as the emergent symmetry of Fermi liquid theory}
\label{sec:loop_group}
\begin{figure}
\includegraphics[scale=0.5]{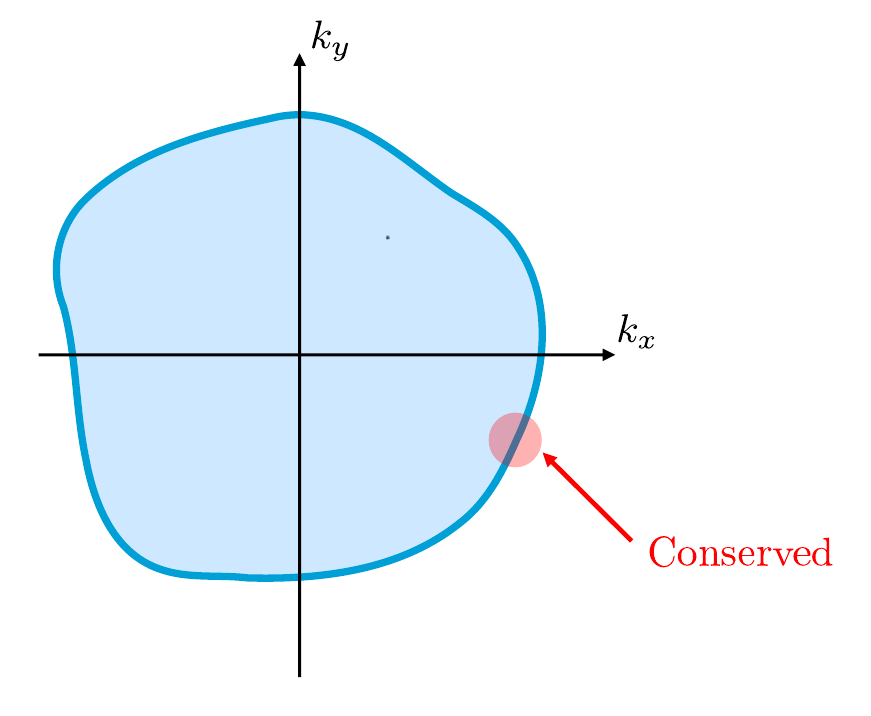}
\caption{\label{fig:fermi_surface}The emergent symmetry of a 2-D Fermi liquid is associated with the conservation of charge at each point of the Fermi surface separately.}
\end{figure}
One of the most basic properties of Fermi liquid theory in spatial dimension $\geq 2$ is that the quasi-particle scattering rate for quasi-particles near the Fermi surface is very slow at low temperatures; specifically, it scales at most like $\sim T^2$. A natural interpretation of this statement is that the number of quasi-particles at every single point on the Fermi surface is an emergent conserved quantity. (What exactly we mean by ``emergent conserved quantity/symmetry'' is actually a subtle point in systems with a Fermi surface -- see the discussion in Section \ref{subsec:conundrum}).

In quantum mechanics, conservation laws exactly correspond to symmetries, so the above considerations immediately imply that we have an emergent symmetry. Moreover, since there are infinitely many independent conserved quantities, the emergent symmetry group is an infinite-dimensional Lie group. Let us now precisely identify what this symmetry group is \cite{Else_2007}. It is sometimes asserted that there is a separate $\mathrm{U}(1)$ symmetry for each point on the Fermi surface. This is not quite right, however: observe that charges of a $\mathrm{U}(1)$ symmetry are quantized to be integers, while there is no meaningful sense in which the charge at a given point on the Fermi surface is quantized. Instead there is a \emph{distribution} $N(\theta)$ on the Fermi surface, which can be integrated over the Fermi surface to give the (quantized) total charge:
\begin{equation}
    Q = \int_{\mathcal{F}} N(\theta) d\theta.
\end{equation}
The integral is over the Fermi surface, which we represent as a compact $(d-1)$-dimensional manifold $\mathcal{F}$. For example, in $d=2$, $\mathcal{F}$ would just be a circle (or a disjoint union of circles if the Fermi surface has multiple components)\footnote{One can also consider slightly more exotic possibilities. For example, at a Lifshitz transition between two different Fermi surface topologies, the Fermi surface is actually a figure-8 space (which is not a manifold).}. All the equations we will write in this review will be independent of the specific parameterization of the Fermi surface.

It also makes physical sense to integrate $N(\theta)$ against any sufficiently regular test function $f(\theta)$:
\begin{equation}
    \label{eq:Q_f}
    Q_f = \int f(\theta) N(\theta) d\theta.
\end{equation}
These will be the infinitesimal generators of the infinite-dimensional Lie group that we are trying to construct.
What we mean by ``sufficiently regular'' is that we should not allow functions $f$ which are too pathological: for example, we certainly would not want to consider functions $f$ which are nowhere continuous!
Because \eqnref{eq:Q_f} corresponds to an integral in momentum space, if $f$ fluctuates too wildly, this will mean that the corresponding quantity $Q_f$ is highly non-local in real space and should not be considered. Precisely which regularity condition one should impose on $f$ may ultimately be determined from physical considerations, but it will not matter very much for the general discussions in this review. However, in order to define the anomaly we will at a minimum require $f$ to be continuous. In what follows, we will refer to ``regular'' functions, leaving the precise definition of ``regularity'' unspecified, except that it will at least imply that the functions are continuous.

\eqnref{eq:Q_f} defines the infinitesimal elements (i.e.\ the Lie algebra) of the infinite-dimensional Lie group. To define the Lie group itself, we have to exponentiate the infinitesimal elements, i.e. we consider
\begin{equation}
    \label{eq:exp_Q_f}
    \exp(i \alpha Q_f) = \exp\left(i \alpha \int_{\mathcal{F}} f(\theta) N(\theta) d\theta\right).
\end{equation}
Observe that since the total charge $Q$ is quantized to integers, we have that if $\alpha f_1(\theta) = \alpha f_2(\theta) + 2\pi n$, where $n$ is an integer, then $\exp(i \alpha Q_{f_1}) = \exp(i \alpha Q_{f_2})$. This naturally leads us to identify operators of the form \eqnref{eq:exp_Q_f} as elements of the infinite-dimensional Abelian Lie group $\LUF$, which is defined to be
\begin{equation}
    \LUF := \{ \mbox{Regular functions $\mathcal{F} \to \mathrm{U}(1)$} \}.
\end{equation}
(with the group operation being pointwise multiplication).
In the case where $\mathcal{F}$ is a circle, we will also simply write $\mathrm{LU}(1)$ for $\LUF$. This is an example of what mathematicians call a ``loop group'' \cite{LoopGroupsBook}.

Note that in general, $\LUF$ also contains elements that are not continuously connected to the identity, and cannot be expressed as the exponential of an infinitesimal generator. For example, if $\mathcal{F}$ is the circle, these correspond to functions $f : S^1 \to \mathrm{U}(1)$ which have non-trivial winding number. In fact, these are also physically allowed symmetry operators. To see this, note that  any continuous function $f : S^1 \to \mathrm{U}(1)$ can be written as $f(\theta) = \exp(2\pi i \mathfrak{f}(\theta))$ where $\mathfrak{f}$ is a real-valued function which may have discontinuities at which it jumps by an integer.
 In general, the problem with having discontinuities in $\mathfrak{f}$ is that $Q_{\mathfrak{f}}$ will then be sensitive to short-distance (in momentum space) fluctuations of the charges on the Fermi surface. If $\mathfrak{f}$ jumps by an amount $\Delta \mathfrak{f}$ at a point $\theta_*$ then a unit charge moving over $\theta_*$ will change $Q_\mathfrak{f}$ by $\pm \Delta \mathfrak{f}$. Therefore, if $\Delta \mathfrak{f}$ is an integer then $Q_\mathfrak{f}$ is still well-defined modulo integers, and so we see that $\exp(2\pi i Q_\mathfrak{f})$ is well-defined.

 In summary, therefore, we have identified the emergent symmetry of Fermi liquid theory in $d$ spatial dimensions: it is $\mathrm{L}_{\mathcal{F}} \mathrm{U}(1)$, where $\mathcal{F}$ is a $(d-1)$-dimensional manifold representing the Fermi surface.

 \section{Anomalies of loop groups}
\label{sec:loop_group_anomalies}
 Now we turn to the question of how to classify the anomalies of the symmetry $\mathrm{L}_{\mathcal{F}} \mathrm{U}(1)$. For simplicity we will mostly the consider the case where $\mathcal{F}$ is a circle so that $\mathrm{L}_{\mathcal{F}} \mathrm{U}(1)= \mathrm{LU}(1)$, as is relevant for 2D Fermi liquids, although the statements can easily be generalized. We will want to use the correspondence between anomalies of a $G$ symmetry in $d$ spatial dimensions and $G$-SPTs in $d+1$ spatial dimensions. From some general abstract arguments invoking the structure that SPT phases are believed to have (see Appendix \ref{appendix:classification}), one can show that there is a map from $\mathrm{U}(1)$ SPTs in $d+2$ spatial dimensions to $\mathrm{LU}(1)$ SPTs in $d+1$ spatial dimensions. Specifically, if we set $d=2$, then $\mathrm{U}(1)$ SPTs in $4$ spatial dimensions have a $\mathbb{Z}$ classification, corresponding to 4D quantum Hall states. These will define  $\mathrm{LU}(1)$ SPTs in $3$ spatial dimensions, which will then descend to $\mathrm{LU}(1)$ anomalies in $2$ spatial dimensions. We will see that this in particular describes the anomaly of Fermi liquid theory in 2 spatial dimensions.

The abstract arguments of Appendix \ref{appendix:classification}, however, do not give much insight into the physical consequences of the anomaly. For this, we would like to write a topological term for the response of an $\mathrm{LU}(1)$ SPT in 3 spatial dimensions, analogous to \eqnref{eq:chern_simons}, which will then allow us to derive an anomaly equation analogous to \eqnref{eq:anomaly_1d} by inflow.

In order to do this, we need to define what we mean by an $\mathrm{LU}(1)$ gauge field. This turns out to be a surprisingly subtle issue. The conventional definition of a gauge field is as follows.
\begin{quote}
    Let $G$ be a Lie group, and let $M$ be a manifold. Then a $G$ gauge field on $M$ is a connection on a principal $G$ bundle over $M$.
\end{quote}
More concretely, with respect to a (local) trivialization of the bundle (i.e.\ a choice of gauge), a $G$ gauge field is a covariant vector field $A_\mu$ on $M$ valued in the Lie algebra of $G$ (where we identify gauge fields differing by a gauge transformation).
In particular, for $G = \mathrm{LU}(1)$, we have a family $A_\mu(\theta)$ of vector fields parameterized by the circle, with gauge transformation
\begin{equation}
    A_\mu(\theta) \to A_\mu(\theta) + \partial_\mu \lambda(\theta),
\end{equation}
where $\lambda(\theta)$ is a scalar function on $M$ that also depends on $\theta$. For reasons that will become clear shortly, we will refer to this as the ``naive'' defition of $\mathrm{LU}(1)$ gauge field.

The problem is that when one tries to study the properties of $\mathrm{LU}(1)$ anomalies, it quickly becomes clear that the naive definition is not the physically correct one, which is related to the fact that $\mathrm{LU}(1)$ is an infinite-dimensional Lie group. For example, there does not appear to be any way to write a topological action for the $\mathrm{LU}(1)$ SPT in 3 spatial dimensions in terms of these naive $\mathrm{LU}(1)$ gauge fields. Instead, Ref.~\cite{Else_2007} argued that a better definition is as follows.
\begin{quote}
    Let $M$ be a manifold. Then an $\mathrm{LU}(1)$ gauge field on $M$ is a $\mathrm{U}(1)$ gauge field on $M \times S^1$, i.e.\ a connection on a principal $\mathrm{U}(1)$ bundle over $M \times S^1$.
\end{quote}
If one unpacks this definition, it does not seem all that different from the naive definition above. However, there is one crucial difference. According to this definition, an $\mathrm{LU}(1)$ gauge field has components $A_\mu(\theta)$ along $M$, and \emph{also} a component $A_\theta(\theta)$ along the circle (i.e. along the Fermi surface). The latter is absent in the naive definition. Moreover, the $A_\theta$ component also transforms under gauge transformations -- the general gauge transformation is
\begin{align}
    A_\mu(\theta) &\to A_\mu(\theta) + \partial_\mu \lambda(\theta). \\
    A_\theta(\theta) &\to A_\theta(\theta) + \partial_\theta \lambda(\theta).
\end{align}

    In a Fermi liquid $A_\theta$ can be interpreted as the quasiparticle Berry connection when quasiparticles are moved around the Fermi surface. But more generally, it would appear that $A_\theta$ should simply be viewed as an essential part of what it means to have an $\mathrm{LU}(1)$ gauge field, in any system with an $\mathrm{LU}(1)$ symmetry. One reason why this makes sense is that $\mathrm{LU}(1)$ symmetry is generally less robust than finite-dimensional Lie group symmetries. For example, it is well known that in Fermi liquid theory, applying a magnetic field will lead to quasiparticles precessing along the Fermi surface, which actually violates the conservation of $\mathrm{LU}(1)$ charge\footnote{Note that this does not (or at least not directly) reflect a quantized anomaly, because the precession rate is determined by non-universal data such as the Fermi velocity and is not quantized.}. If quasiparticles move along the Fermi surface, then they can feel a ``gauge field'' as they move, which is the quasiparticle Berry connection.

    With the above definition of $\mathrm{LU}(1)$ gauge field, it is now clear how to write a topological term describing a 3D $\mathrm{LU}(1)$ SPT. Specifically, for a gauge field $A$ on a 4-dimensional space-time $M$, we write the action \cite{Else_2007}
    \begin{equation}
        S[A] = \frac{m}{24\pi^2} \int_{M \times S^1} A \wedge dA \wedge dA,
    \end{equation}
    where we view $A$ as $\mathrm{U}(1)$ gauge field on the 5D manifold $M \times S^1$, in order to write the 5D Chern-Simons term. This is the same Chern-Simons term that would describe a $\mathrm{U}(1)$ SPT in 4 spatial dimensions, i.e. a 4D Quantum Hall state. By inflow, one then obtains the anomaly equation
    \begin{equation}
        \label{eq:anomaly}
        \partial_\mu j^\mu = \frac{m}{8\pi^2}\epsilon^{\lambda \sigma \tau \kappa} (\partial_\lambda A_\sigma) (\partial_\tau A_\kappa).
    \end{equation}
    Here we take our indices $\mu$, etc.\ to range not only over the 3 space-time indices of a system in 2 spatial dimensions, but also over the $\theta$ direction. Moreover, all the quantities can depend on $\theta$ as well as the space-time coordinate. The anomaly coefficient $m$ is quantized to be an integer, and it will turn out that it takes the value $\pm 1$ in a spinless Fermi liquid (the sign depends on an arbitrary choice of orientation for the Fermi surface).

    In Section \ref{sec:consequences}, we will explore the consequences of the anomaly. However, first we will discuss whether we should expect the considerations described here to apply also to non-Fermi liquids, not just Fermi liquids.
    
    \section{Beyond Fermi liquids: compressibility and ersatz Fermi liquids}
    \label{sec:beyond_fermi}
    As shown in Section \ref{sec:consequences} below, we can rederive various properties of Fermi liquid theory from the anomaly perspective. However, the real power comes by applying the ideas to non-Fermi liquid metals. In general, such metals are very challenging to understand analytically. However, the anomaly is an exact, non-perturbative statement and all the properties described in Section \ref{sec:consequences} will immediately follow in \emph{any} system that has an emergent $\mathrm{LU}(1)$ symmetry with nonzero anomaly coefficient. Ref.~\cite{Else_2007} coined the term ``ersatz Fermi liquid'' for such a system.
    
    The salient question is: is \emph{every} non-Fermi liquid metal an ersatz Fermi liquid? Ref.~\cite{Else_2007} conjectured that the answer is yes, assuming that one restricts to systems that have a microsopic lattice translation symmetry (in other words, we consider clean systems without impurities). Rigorous proofs have not been given, but Ref.~\cite{Else_2007} did give the following argument. Both Fermi liquids and non-Fermi liquid metals generally have the property that they are \emph{compressible}. Specifically this means that in a system with microscopic lattice translation symmetry and $\mathrm{U}(1)$ charge conservation, the \emph{filling} $\nu$ (charge per unit cell) can be tuned continuously by varying microscopic parameters of the Hamiltonian, and in particular can take irrational values. Ref.~\cite{Else_2007} considered the matching conditions between the microscopic filling and the anomaly of the emergent symmetry (generalizing Luttinger's theorem, see for example Section \ref{subsec:projective_magnetic}), and showed that if $\nu$ is irrational, then the emergent symmetry group cannot be a compact finite-dimensional Lie group. (Specifically this refers to the \emph{internal} emergent symmetries, i.e. those which do not move points in space-time around). Since non-compact finite-dimensional Lie group internal symmetries do not seem very physical, this leaves only the possibility of an infinite-dimensional symmetry, for which the $\mathrm{LU}(1)$ symmetry that occurs in Fermi liquid theory seems by far the most natural candidate. It is a very interesting question for the future to determine if there are any other possibilities. (Another state that is also compatible with compressibility, albeit one that would not normally be referred to as a ``metal'', is a superfluid, in which the compressibility is instead related to an emergent higher-form symmetry \cite{Delacretaz_1908,Else_2106}).

Finally, let us briefly mention how the emergent conservation law arises in systems of fermions coupled to a critically fluctuating boson \cite{Hertz__1976,Millis__1993}. The fermions can scatter off the boson and move from one point on the Fermi surface to another. However, at low energies the fluctuations of the boson occur near $\mathbf{k} = 0$, and therefore the scattering can only connect nearby points on the Fermi surface. The lower the energy scale, the less distance on the Fermi surface the fermions can scatter over. This is consistent with the idea of an emergent $\mathrm{LU}(1)$ symmetry. (One can also have bosons with critical fluctuations near another point in momentum space, e.g.\ if the boson represents an order parameter for anti-ferromagnetic or charge-density-wave order, but in that case the boson generally only connects a finite number of ``hot spots'' on the Fermi surface, leaving the emergent conservation law at all other points on the Fermi surface unaffected.) Finally, we should note that when the boson is odd under time-reveral and inversion symmetry, one can find an RG fixed point with a nonzero value of the BCS interaction \cite{Metlitski_1403,Raghu_1507}. In this case, the $\mathrm{LU}(1)$ symmetry is broken down to an infinite-dimensional subgroup \cite{Shi_2204}; specifically, this subgroup corresponds to those functions $ f : S^1 \to \mathrm{U}(1)$ which are either constant or odd under inversion symmetry. Most of the considerations of this review will still apply when this subgroup is the emergent symmetry group rather than the full $\mathrm{LU}(1)$.

    \section{Consequences of the emergent symmetry and anomaly} \label{sec:consequences}

    In this section, we will consider various quantities which can be derived directly from the emergent symmetry and anomaly. Therefore, they will hold in any ersatz Fermi liquid.
    
    \subsection{Non-conservation of $\mathrm{LU}(1)$ charges in response to an electric field}
    Suppose we apply an electric field $E_i$ to the system, which corresponds to taking $\partial_t A_i - \partial_i A_t = E_i$. In Fermi liquid theory, one can show that the Fermi surface charges $N(\theta)$ will become non-conserved in response to the electric field. One can view this as a manifestation to the anomaly \cite{Else_2007}. Specifically, in Fermi liquid theory one has
    \begin{equation}
        \label{eq:fl_anomaly}
        \partial_\mu j^\mu(\theta) = \frac{1}{(2\pi)^2} \epsilon^{ij} E_i \partial_\theta k_j(\theta),
    \end{equation}
    where $j^\mu(\theta)$ is the current associated to the charge $N(\theta)$.
    and $k_i(\theta)$ is the momentum of the point on the Fermi surface labeled by $\theta$.
    
    In order to match with the general anomaly equation \eqnref{eq:anomaly}, it is clear that we need to take the anomaly coefficient $m=1$ and set
    \begin{equation}
        \label{eq:phase_space_curvature}
    \partial_\theta A_i(\theta) - \partial_i A_\theta = \partial_\theta k_i.
    \end{equation}
    A nice interpretation of \eqnref{eq:phase_space_curvature} was given in \cite{Lu_2302}: it essentially encodes the fact that quasiparticles see position and momentum as non-commuting coordinates. \eqnref{eq:phase_space_curvature} should continue to hold even in non-Fermi liquids, and thus \eqnref{eq:fl_anomaly} will hold in any ersatz Fermi liquid.

    \subsection{Projective representation of $\mathrm{LU}(1)$ in response to magnetic field}
    \label{subsec:projective_magnetic}
    Suppose that instead of an electric field, we apply a (possibly spatially dependent) magnetic field $B(\mathbf{x})$. Ref.\cite{Else_2007} showed that from the anomaly equation, one can derive that the $\mathrm{LU}(1)$ symmetry acts projectively in the presence of this magnetic field: in particular, while the $\mathrm{LU}(1)$ symmetry is Abelian, the unitary operators implementing the $\mathrm{LU}(1)$ symmetry in the presence of a magnetic field commute only up to a phase:
    \begin{equation}
        \label{eq:magnetic_field_commutator}
        U_f U_g U_f^{\dagger} U_g^{\dagger} = \exp\left(\frac{im\Phi}{(2\pi)^2} \int f(\theta) g'(\theta) d\theta\right),
    \end{equation}
    where $f , g : S^1 \to \mathrm{U}(1)$ describe elements of $\mathrm{LU}(1)$.
    where $\Phi = \int B(\mathbf{x}) d^2 \mathbf{x}$ is the total magnetic flux (which must be an integer multiple of $2\pi$).
    One can also express this in terms of the infinitesimal generators as 
    \begin{equation}
        \label{eq:Ntheta_commutator}
    [N(\theta), N(\theta')] = \frac{im\Phi}{(2\pi)^2} \delta'(\theta - \theta')
    \end{equation}
    This result was already known in Fermi liquid theory \cite{Golkar_1602,Barci_1805}; for example, Ref.~\cite{Else_2007} showed that in the semiclassical theory of electron transport, one can derive the Poisson bracket version of \eqnref{eq:Ntheta_commutator}.

    One consequence of \eqnref{eq:magnetic_field_commutator} is that it allows for an elegant proof \cite{Else_2007} of Luttinger's theorem for 2D Fermi liquids, which relates the microscopic charge filling $\nu$ (the charge per unit cell of translations) to the volume enclosed by the Fermi surface. (Hence, Luttinger's theorem is automatically extended to any 2D ersatz Fermi liquid). The point is that when the filling $\nu$ is not an integer, the lattice translation symmetries $\mathbb{T}_x$ and $\mathbb{T}_y$ must become non-commutative in the presence of a background magnetic flux $\Phi$. Specifically, we have
    \begin{equation}
        \label{eq:filling_anomaly}
    \mathbb{T}_x \mathbb{T}_y \mathbb{T}_x^\dagger \mathbb{T}_y^\dagger = e^{i \Phi \nu},
    \end{equation}
    In the low-energy effective theory, these symmetries must map to elements of $\mathrm{LU}(1)$, corresponding to functions $k_x, k_y : S^1 \to \mathrm{LU}(1)$. The functions $k_x, k_y$ trace out a closed curve in momentum space (i.e. the Brillouin zone), which we can interpret as the Fermi surface. Then equating \eqnref{eq:magnetic_field_commutator} with \eqnref{eq:filling_anomaly} gives
    \begin{equation}
    \nu = m \frac{\mathcal{V}_F}{(2\pi)^2} \quad \mbox{(mod 1)}
    \end{equation}
    where
    \begin{equation}
    \mathcal{V}_F = \int k_x(\theta) k_y'(\theta) d\theta
    \end{equation}
    is the volume enclosed by the Fermi surface. This is precisely Luttinger's theorem.

        \subsection{Quantum oscillations}
        A well known feature of Fermi liquid theory is that certain physical properties, such as the conductivity (Shubnikov--de Haas oscillations)  and, in spinful systems, the magnetic susceptibility (de Haas--van Alphen oscillations), vary periodically in $1/B$, where $B$ is the applied magnetic field \cite{QuantumOscillationsBook}. In two spatial dimensions, the periodicity is given by
        \begin{equation}
            \label{eq:quantum_oscillations}
            \Delta(1/B) = \frac{2\pi}{\mathcal{V}_F},
        \end{equation}
        where $\mathcal{V}_F$ is the volume enclosed by the Fermi surface. In three dimensions, the periodicity is also given by \eqnref{eq:quantum_oscillations}, with $\mathcal{V}_F$ replaced by the extremal cross-sectional area of the Fermi surface in the plane perpendicular to the magnetic field.

        The conventional argument for \eqnref{eq:quantum_oscillations} involves tracking quasiparticles that precess around the Fermi surface in the presence of a magnetic field. However, much more generally, the periodicity can simply be viewed as the manifestation of an emergent \emph{continuous} translation symmetry \cite{Else_2007} with effective charge density
        \begin{equation}
            \label{eq:effective_charge_density}
            \rho = \frac{1}{(2\pi)^2} \mathcal{V}_F.
        \end{equation}
        The effective charge density can be defined, for example, through the non-conservation of the momentum in the presence of an electric field:
        \begin{equation}
            \partial_\mu T\indices{^\mu_i} = \rho E_i.
        \end{equation}
        where $T\indices{^\mu_i}$ is the current of the momentum $P_i$.

        In particular, one can show\footnote{Ref.~\cite{Else_2007} gave a somewhat circuitous argument involving an appeal to a UV completion; however, one can actually derive it directly in the IR theory.} that such an emergent symmetry is present, with the effective charge density \eqnref{eq:effective_charge_density} in any ersatz Fermi liquid with the property that the Fermi surface does not wrap non-trivially around the Brillouin zone (if the anomaly coefficient $m$ is not equal to 1, then the right-hand side of \eqnref{eq:effective_charge_density} gets multiplied by $m$). Therefore, any 2D ersatz Fermi liquid should exhibit quantum oscillations, regardless of the existence of quasiparticles.

        Ref.~\cite{Else_2007} also gave arguments suggesting that 3D ersatz Fermi liquids should also exhibit quantum oscillations satisfying \eqnref{eq:effective_charge_density} with $\mathcal{V}_F$ defined as the extremal cross-section. These arguments involved compactifying one spatial dimension to reduce to a 2D system and making a reasonable assumption about the relation between the 3D Fermi surface and the collection of 2D Fermi surfaces obtained upon such dimensional reduction. It would still be interesting to have a purely 3D argument for this result.

        \subsection{Optical conductivity and Drude weight}
        \label{subsec:drude_weight}
        One feature of metals is that they have very low electrical resistivity; in fact, in a clean metal the DC resistivity goes to zero at temperature $T=0$. One can view this as a consequence of the emergent symmetry and anomaly. Let us consider a system with an $\mathrm{LU}(1)$ symmetry with nonzero anomaly coefficient $m$ (for the moment we treat the $\mathrm{LU}(1)$ symmetry as exact). We define the ``Drude weight'' to be the coefficient of the pole at $\omega = 0$ in the optical conductivity:
        \begin{equation}
         \sigma(\omega) = \mathcal{D} \frac{i}{\omega} + \cdots.
        \end{equation}
        In particular, if $\mathcal{D}$ is nonzero it means that the DC resistivity is exactly zero.

        One can show \cite{Else_2010,Else_2106} that the Drude weight is actually fixed by the anomaly to a form that only depends on the shape of the Fermi surface and the susceptibilities of the conserved quantities associated with the $\mathrm{LU}(1)$ symmetry. Specifically, one finds
        \begin{equation}
            \label{eq:drude_weight}
            \mathcal{D}^{ij} = \frac{m^2}{(2\pi)^4}\int d\theta \int d\theta' \chi^{-1}(\theta,\theta') w^i(\theta) w^j(\theta')
        \end{equation}
       Here we have introduced the inverse susceptibility matrix $\chi^{-1}(\theta,\theta')$, which is a thermodynamic quantity that can be defined, for example, by
        \begin{equation}
            \label{eq:susceptibility_matrix}
            \chi^{-1}(\theta,\theta') = \frac{\delta^2 \varepsilon}{\delta n(\theta) \delta n(\theta')},
        \end{equation}
        where $\varepsilon$ is the energy density, and $n(\theta)$ is the density of the conserved charge $N(\theta)$. (Note that strictly speaking the derivatives should be computed holding the entropy density fixed \cite{Else_2301}; at $T=0$ this is equivalent to computing the derivatives with temperature held fixed.) We also have defined
        \begin{equation}
            w^i(\theta) = \epsilon^{ij} \partial_\theta k_j(\theta),
        \end{equation}
        where $k_i(\theta)$ is the (vector) Fermi momentum at the point of the Fermi surface specified by $\theta$.
        
        In Fermi liquid theory, $\chi^{-1}$ can be computed in terms of the Fermi velocity and the Landau parameters, and one can show that \eqnref{eq:drude_weight} reduces to the standard formla for the Drude weight of a Fermi liquid. More generally, since thermodynamic stability requires that the inverse susceptibility matrix be positive-semidefinite, \eqnref{eq:drude_weight} is generally a positive-semidefinite matrix. In principle, it is possible for it to be zero \cite{Else_2007,Else_2106}, but this generally requires fine-tuning infinitely many components of the susceptibility matrix \cite{Shi_2208}.

        The fact that the $\mathrm{LU}(1)$ symmetry is emergent means that there can be irrelevant operators that violate it. The consequence of this is that one expects the optical conductivity at nonzero temperature to take the form
        \begin{equation}
            \label{eq:optical_conductivity}
            \sigma(\omega) = \frac{\mathcal{D}}{i \omega + \gamma(T)}.
        \end{equation}
        where $\mathcal{D}$ is still given by \eqnref{eq:drude_weight} (which in principle could be temperature-dependent), and the ``scattering rate'' $\gamma(T)$ (more precisely, the rate of violation of the $\mathrm{LU}(1)$ conservation law) should go to zero sufficiently fast as $T \to 0$ due to its origin in irrelevant operators. (For a more careful discussion of what ``sufficiently fast'' means, see Section \ref{subsec:conundrum}). For example, in Fermi liquid theory, $\gamma(T)$ arises from umklapp scattering and scales at most like $\sim T^2$.

        As a side note, we observe that the zero resistivity in a superfluid can also be viewed \cite{Else_2106} as a consequence of an emergent symmetry with an anomaly, specifically a $(d-1)$-form symmetry in $d$ spatial dimensions (this symmetry follows from the confinement of vortices in a superfluid \cite{Delacretaz_1908}), which is an example of a ``generalized global symmetry'' \cite{Gaiotto_1412}. The fact that that the zero resistivity is much more robust in a superfluid than a metal can be attributed to the fact that higher-form symmetries are much more robust than conventional global symmetries (i.e.\ ``0-form symmetries''), and cannot be broken even by irrelevant operators.

        \subsection{(Hydro)dynamics}
        In this section we will consider the dynamical behavior of the system; that is, the oscillation modes of the system at non-zero frequency $\omega$ and wave-vector $\mathbf{q}$. Since there is an emergent $\mathrm{LU}(1)$ symmetry, for small enough $\omega$ and $q$, it should be a good approximation to treat this as an exact symmetry. The resulting infinitely many conservation laws will strongly constrain the dynamics. Let us also suppose that we can treat the system from the point of view of hydrodynamics. In hydrodynamics, one assumes that the system is approximately in \emph{local} thermal equilibrium at all points in space and time. The system is then characterized by the space- and time-dependent values of the thermodynamic quantities (temperature, chemical potential, etc.) which characterize the local thermal equilibrium state. These evolve according to hydrodynamic equations of motion which can be written in terms of a gradient expansion. The thing to keep in mind is that if we treat the infinitely many conservation laws arising from the $\mathrm{LU}(1)$ symmetry as exact, the local equilibrium state will be characterized by infinitely many thermodynamic variables: that is, the ``chemical potential'' associated with each of the conserved quantities. Therefore, the hydrodynamic equations will describe the evolution of infinitely many quantities at each point in space-time.

        This should not be too surprising, because the known dynamics of Fermi liquid theory in the collisionless regime also involves infinitely many quantities at each point in space-time, specifying the shape of the local Fermi surface. In fact, one can show \cite{Else_2301} that the linearized version of the equations of motion of Fermi liquid theory exactly agrees with what one would obtain from the hydrodynamic equations of motion, truncated to zero-th order in the gradient expansion.

        More generally, as shown in Ref.~\cite{Else_2301}, the linearized hydrodynamic equations of motion at zero-th order in the gradient expansion are actually fully constrained by the $\mathrm{LU}(1)$ symmetry and anomaly (see also Ref.~\cite{Huang_2402} for a non-equilibrium effective field theory perspective). The only parameters are the thermodynamic susceptibilities defined by \eqnref{eq:susceptibility_matrix}. Therefore, in any ersatz Fermi liquid, one expects the oscillation mode spectrum to be essentially the same as in a Fermi liquid, including zero sound modes (depending on the specific paramater values) and a particle-hole continuum.

        It is worth remarking that these results depend on the assumption that the system is in local thermal equilibrium. In particular, hydrodynamics cannot be expected to describe the behavior of the system at frequencies $\omega$ larger than the inverse of the thermalization time. In particular, the concern would be that at low temperatures $T$, the thermalization time would be very large, perhaps at the minimum $\sim T^{-1}$ \cite{Delacretaz_2310}. That would suggest that the hydrodynamics might only capture the response for $\omega \ll T$, and in particular might not be valid at all at $T=0$. On the other hand, we do know that in Fermi liquid theory the Boltzmann equation is valid regardless of the ratio $\omega/T$, and in particular remains valid at $T=0$. Whether this is also true in non-Fermi liquids remains unclear \cite{Kim_9504,Else_2301}.

\section{Limitations and extensions}
\label{sec:limitations}

    \subsection{Nonlinear responses and violation of the $\mathrm{LU}(1)$ symmetry.}
    \label{subsec:nonlinear}
        All the manifestations of the anomaly discusonesed in Section \ref{sec:consequences} can essentially be viewed as \emph{linear} responses.  Beyond linear responses, although one could try to formally compute things using the $\mathrm{LU}(1)$ anomaly equation, the problem is that even in Fermi liquid theory, the $\mathrm{LU}(1)$ symmetry is violated beyond linear order in perturbations about the ground state, even in the absence of any external electric or magnetic fields (so the violation cannot simply attributed to the anomaly, as the right-hand side of the anomaly equation \eqnref{eq:anomaly} would be zero in this case).

        This fact was previously pointed out in the context of the nonlinear bosonization framework of Ref.~\cite{Delacretaz_2203}. Here I will give a  more elementary argument. Recall that in Fermi liquid theory, if one ignores quasiparticle scattering then the time-evolution of the quasiparticle distribution $f(\mathbf{p},\mathbf{x},t)$ in the presence of an external electric field $\mathbf{E}$, is governed by the Boltzmann equation
    \begin{equation}
        \label{eq:boltzmann}
    \frac{\partial f}{\partial t} + \left(\frac{\partial \epsilon}{\partial \mathbf{x}}  + \mathbf{E}\right) \cdot \frac{\partial f}{\partial \mathbf{p}} - \frac{\partial \epsilon}{\partial \mathbf{p}} \cdot \frac{\partial f}{\partial \mathbf{x}} = 0,
    \end{equation}
    where the quasiparticle energy $\epsilon(\mathbf{p},\mathbf{x},t)$ is given by
    \begin{equation}
    \epsilon(\mathbf{p},\mathbf{x},t) = \epsilon_0(\mathbf{p}) + \int V(\mathbf{p},\mathbf{p}') \delta f(\mathbf{p}',\mathbf{x},t),
    \end{equation}
    where $\delta f = f - f_0$, with $f_0$ the equilibrium distribution in the ground state, and
    and $V(\mathbf{p},\mathbf{p}')$ denotes the Landau interaction.
    From \eqnref{eq:boltzmann} one can, for example, derive \eqnref{eq:fl_anomaly} at linear order in the electric field, showing that the Fermi surface charges $N(\theta)$ become non-conserved in the presence of an electric field, reflecting the 't Hooft anomaly \eqnref{eq:anomaly}. The problem is that we see from \eqnref{eq:boltzmann} that a spatial gradient of the electron distribution function $f$ itself contributes an effective ``electric field'' and therefore will lead to non-conservation of $N(\theta)$. However, because no actual external electric field is being applied, the right-hand side of the anomaly equation \eqnref{eq:anomaly} would be zero. Note that this effect enters only at order $(\delta f)^2$, because we can rewrite the offending term of \eqnref{eq:boltzmann} as
    \begin{align}
    \frac{\partial \epsilon}{\partial \mathbf{x}} \cdot \frac{\partial f}{\partial \mathbf{p}} &= \frac{\partial \delta \epsilon}{\partial \mathbf{x}} \cdot \frac{\partial f}{\partial \mathbf{p}} \\ &=
    \frac{\partial}{\partial \mathbf{x}} \cdot \left(\delta \epsilon \frac{\partial f}{\partial \mathbf{p}}\right) - \delta \epsilon \frac{\partial}{\partial \mathbf{x}} \cdot \frac{\partial f}{\partial \mathbf{p}} \\
    &= \frac{\partial}{\partial \mathbf{x}} \cdot \left(\epsilon \frac{\partial f}{\partial \mathbf{p}}\right) - \delta \epsilon \frac{\partial}{\partial \mathbf{x}} \cdot \frac{\partial \delta f}{\partial \mathbf{p}}.
    \end{align}
    where we defined $\delta \epsilon = \epsilon - \epsilon_0$, and used the fact that $f_0$ and $\epsilon_0$ do not depend on $\mathbf{x}$.
    The first term in the right-side does not lead to any non-conservation, since it goes to zero when integrated over space, while the second term is order $O(\delta f)^2$.

    The fact that $\mathrm{LU}(1)$ symmetry is violated in non-linear responses does not necessarily contradict the statement that there is an emergent $\mathrm{LU}(1)$ symmetry, since we can view the non-linear responses as being related to irrelevant operators (however, see also the discussion in Section \ref{subsec:conundrum} regarding the difficulty of defining ``emergent symmetry'' in systems with a Fermi surface). Nevertheless, it would be nice to have a generalization of the $\mathrm{LU}(1)$ anomaly story that can capture non-linear responses. This may be related to the ``non-linear bosonization'' framework for collisionless Fermi liquid theory presented in Ref.~\cite{Delacretaz_2203}. It is worth noting, however, that that framework is specifically designed to describe Fermi liquid theory, whereas the $\mathrm{LU}(1)$ emergent symmetry and anomaly are expected to apply very generally to non-Fermi liquids as well, as we mentioned above.

    \subsection{The conundrum of $T$-linear resistivity in strange metals}
    \label{subsec:conundrum}
    A particular challenge is in understanding how the above general considerations are compatible with the experimental observations of resistivity that is \emph{linear} in temperature $T$ in materials such as high-$T_c$ cuprates \cite{Proust_1807,Varma_1908}. Let us hypothesize that the $T$-linear resistivity can occur even in a perfectly clean system that satisfies compressibility (thus, we do not consider models such as those proposed in Ref.~\cite{Patel_2203} in which the resistivity is inherently linked to impurities.). Then from the discussion in Section \ref{sec:beyond_fermi} we should expect an emergent $\mathrm{LU}(1)$ symmetry, and we can invoke the discussion of optical conductivity and Drude weight described in Section \ref{subsec:drude_weight}.

    Let us additionally assume that the resistivity is actually linear in $T$ all the way down to zero temperature once superconductivity is suppressed. 
    Finally, let us suppose that the Drude weight predicted by \eqnref{eq:drude_weight} remains nonzero as $T \to 0$. (Thus, we are discounting the ``critical drag'' mechanism proposed in Refs.~\cite{Else_2010,Else_2106}, which may require fine-tuning \cite{Shi_2208}). Then a $T$-linear DC resistivity suggests that the ``scattering rate'' $\gamma(T)$, i.e.\ the rate of violation of $\mathrm{LU}(1)$ conservation, that appears in \eqnref{eq:optical_conductivity} is linear in $T$.

    Is this compatible with the idea of an emergent $\mathrm{LU}(1)$ symmetry? Conventionally, what emergent symmetry means is that we have an RG fixed point, such that any operator that breaks the symmetry is irrelevant. In that case, one can use scaling arguments to show \cite{Else_2010} that $\gamma(T)$ should scale as $\sim T^{1 + \epsilon}$ for some $\epsilon > 0$ (or it could be something like $\sim T \log T$ if there are marginally irrelevant operators). Thus, a $T$-linear scattering rate seems incompatible with an emergent $\mathrm{LU}(1)$ symmetry.

    The problem with these arguments is that the sense in which a system with a Fermi surface can be described by an RG fixed point is subtle. Normally one expects that RG fixed points should be scale-invariant, and yet Fermi surfaces come with an intrinsic scale: the Fermi momentum $k_F$. Treatments such as Refs.~\cite{Polchinski_9210,Shankar_9307} attempt to get around this in Fermi liquid theory by defining the RG flow such that it rescales momentum towards the Fermi surface instead of towards $k=0$. However, this may amount to brushing the problem under the rug, since the ratio $\Lambda/k_F$, where $\Lambda$ is the UV cutoff, still ends up playing a role in these treatments, reflecting a lack of scale invariance. Instead, it may make more sense to view the Fermi liquid as representing an RG \emph{trajectory} rather than a fixed point \cite{Ma_2302,Kukreja_2405}.
    
    For these reasons, it is difficult to give a precise definition of ``emergent symmetry'' in systems with Fermi surfaces. Instead, we could ask a more operational question: what $T$-dependence of the ``scattering rate'' $\gamma(T)$ defined by \eqnref{eq:optical_conductivity} is allowed in a clean compressible system? If $\gamma(T)$ at least goes to zero as $T \to 0$ (which includes the case of $T$-linearity), then one might hope that the $\mathrm{LU}(1)$ ``symmetry'' is still a useful tool to understand the low-energy physics, whether or not we can view it as an emergent symmetry in any precisely defined sense. These are profound and important questions which I hope will be addressed in future work.
    
    \subsection{Fermi surface anomalies and symmetric mass generation}
    In this section, I will briefly describe an alternative perspective on Fermi surface anomalies proposed in Ref.~\cite{Lu_2302}. Recall from the discussion above that the anomaly of $\mathrm{LU}(1)$ descends from an anomaly of $\mathrm{U}(1)$ in one higher dimension -- or, more generally, in $d$ spatial dimensions where the Fermi surface $\mathcal{F}$ is a $d-1$-dimensional manifold, the anomaly of $\mathrm{L}_{\mathcal{F}}\mathrm{U}(1)$ descends from an anomaly of $\mathrm{U}(1)$ in $2d+1$ spatial dimensions. By contrast, Ref.~\cite{Lu_2302} proposed that in any spatial dimension, the ``anomaly'' of a Fermi surface should be classified by $\mathrm{U}(1)$ SPT phases in \emph{zero} spatial dimensions. Note that, in either case, this gives a $\mathbb{Z}$ classification of anomalies, so the two perspectives need not be viewed as contradictory.

    However, the power of the approach described in Ref.~\cite{Lu_2302} seems to be in characterizing when it is possible to gap out the Fermi surface by adding symmetry-breaking terms to the Hamiltonian. Consider for example, starting from a Fermi liquid in $d=2$ and adding a superconducting pairing terms that breaks $\mathrm{U}(1)$ charge conservation down to $\mathbb{Z}_2^f$ (fermion parity). For a single-component (spinless) Fermi surface, $s$-wave pairing is not possible, while $p$-wave pairing does not completely gap out the Fermi surface. To completely gap out the Fermi surface, one can add a $p_x + i p_y$ pairing term, but the gapped ground state is now in a non-trivial topological superconductor phase. By contrast, for a spinful Fermi liquid, $s$-wave pairing is possible in the spin singlet channel, leading to a trivial gapped superconductor.

    One can analyze this using the perspective of Ref.~\cite{Lu_2302}. Specifically, when breaking $\mathrm{U}(1)$ symmetry down to $\mathbb{Z}_2^f$, the $\mathbb{Z}$ classification of 0-D SPTs collapses down to $\mathbb{Z}_2$. Therefore, the ``anomaly'' is non-trivial in the case of a single-component Fermi surface, but for two copies of the system, i.e.\ a spinful Fermi liquid, the ``anomaly'' is trivialized. Ref.~\cite{Lu_2302} also considered scenarios in which the symmetry is such that interactions (i.e. not just free-fermion terms in the Hamiltonian which are quadratic in the fermion operators) are required to gap out the Fermi surface, which again is allowed only when the ``anomaly'' is -trivial\footnote{Note that there is some subtlety in what we mean by ``gapping out the Fermi surface''. Once we break $\mathrm{U}(1)$ charge conservation there is no concept of filling, hence there is no Luttinger's theorem that prevents one from deforming the volume enclosed by the Fermi surface to zero and thus eliminating it. To avoid this, roughly  speaking one should demand that the interactions only affect the degrees of freedom near the Fermi surface.}. These cases were referred to as ``Fermi surface symmetric mass generation''.

    Instead of repeating the arguments of Ref.~\cite{Lu_2302}, here I will give an alternative perspective on the anomaly classification they proposed. Consider a system in $d$ spatial dimensions with $\mathrm{U}(1)$ charge conservation and lattice translation symmetry. Imagine starting from an insulator, nucleating a very small Fermi surface in the Brillouin zone, then expanding the Fermi surface until the volume enclosed fills the entire Brillouin zone. At that point we have another insulator, but it differs from from the one we started with: the filling (charge per unit cell) has increased by 1. We can view this as stacking an SPT protected by $\mathrm{U}(1)$ and lattice translation symmetry. In general, for any internal symmetry $G$, in $d$ spatial dimensions the SPTs protected by $G$ and lattice translation symmetry are in one-to-one correspondence with  $G$ SPTs in $d=0$, thus recovering the classification proposed in Ref.~\cite{Lu_2302}. We can roughly see why, if one obtains a non-trivial SPT via the process described, it should represent an obstruction to ``trivially'' gapping out the Fermi surface; if this were possible, then the entire process would be a continuous deformation of gapped states, but it should not be possible to deform two different SPT phases into each other by such a process.

    Note that in the general case, it does not seem possible to interpret the ``anomaly'' of Ref.~\cite{Lu_2302} as a 't Hooft anomaly of a symmetry (emergent or not) in the conventional sense. For example, the pairing terms which break $\mathrm{U}(1)$ down to $\mathbb{Z}_2^f$ also break the $\mathrm{LU}(1)$ symmetry, so that $\mathbb{Z}_2^f$ is the only internal symmetry remaining, and there are no non-trivial 't Hooft anomalies for systems with only fermion parity symmetry in $d=2$. It will be interesting in the future to understand these ``anomalies'' better and determine whether they have any other physical consequences, beyond representing an obstruction to trivially gapping out the Fermi surface.

    \section{Beyond metals}
    \label{sec:beyond_metals}
Identifying the emergent symmetry and anomaly is a powerful tool in any quantum many-body system, not just metals. Here I will briefly discuss some examples of non-metallic systems in which the emergent symmetry and anomaly structure is very closely analogous to the one for metals.

The first class \cite{Hughes_2401} involves systems in which the gapless surface in momentum space has a different dimensionality $k$ compared to a Fermi surface (in which we always have $k = d-1$, where $d$ is the spatial dimension). One can consider, for example, nodal line semimetals, in which $k=1$. Semi-metals which have only gapless points in momentum space correspond to the case $k=0$. In general, one expects the emergent symmetry in such systems to be $L_{\mathcal{N}} \mathrm{U}(1)$, where $\mathcal{N}$ represents the $k$-dimensional gapless surface in momentum space. One can write an analog of the topological terms described in this review whenever $d+k$ is odd. For $k < d-1$ these are precisely the cases in which there exist stable nodal-surface semi-metals in the non-interacting case, which are protected by a non-trivial Berry Chern number for the bands enclosing the gapless surface. In general if there are additional symmetries such as crystalline symmetries, it is possible to stabilize nodal-surface semi-metals for $d+k$ even as well, and these can also be interpreted in terms of anomalies taking into account the additional symmetry.

The second class \cite{Else_2007} involves cases where there is a Fermi surface of emergent quasiparticles such as a spinon Fermi surface in a spin liquid, or ``composite fermions'' in the ``composite Fermi liquid'' picture of the fractional quantum Hall effect at magnetic filling $\nu = 1/2$. In both cases, the emergent symmetry group is actually a central extension of the loop group. The reason is in these systems, the magnetic flux is not a background, but actually a dynamical quantity, so \eqnref{eq:magnetic_field_commutator} becomes a statement of the non-commutativity of the symmetry group even without any background fields. In these cases the anomaly of the emergent symmetry group is actually $\mathbb{Z}_2$ classified rather than $\mathbb{Z}$ classified, and the composite Fermi liquid realizes the non-trivial anomaly class. Intriguingly, the precise form of the emergent symmetry group seems to depend on whether one uses the Halperin-Lee-Read (HLR) \cite{Halperin__1993} or Son \cite{Son_1502} theories of the composite Fermi liquid, suggesting a subtle difference between these two theories.

\section{Outlook}
\label{sec:outlook}
The main result that we have presented in this paper is that a number of properties of Fermi liquid metals can be understood in terms of the emergent symmetry and anomaly, and therefore can be extended non-perturbatively to non-Fermi liquid metals as well. One objection that could be raised is that since the emergent symmetry and anomaly are shared between Fermi liquids and non-Fermi liquids, they do not immediately allow us to access the properties that are unique to non-Fermi liquids, such as unconventional transport behavior (including the famous $T$-linear resistivity of strange metals). One main question for the future is how to use the improved structural understanding of non-Fermi liquids as a starting point to develop better models or better analyze existing models. For example, the results of Ref.~\cite{Shi_2204,Shi_2208,Shi_2402} deriving exact results for certain models of metallic quantum critical points were partly inspired by the emergent symmetry/anomaly perspective. Another direction was described in Ref.~\cite{Else_2307}, in which a new holographic model of a non-Fermi liquid was constructed that directly builds in the emergent symmetry and anomaly.

Understanding metals, and specifically non-Fermi liquid metals, remains one of the deepest outstanding questions in theoretical condensed matter physics. There is every reason to believe that the existing models are only scratching the surface of the theoretical possibilities. Given the generality of the results described here, the hope is that they can serve as a reference point as we continue to explore this uncharted sea.

\section*{Acknowledgments}
I thank T. Senthil, Zhengyan Darius Shi, John McGreevy, Sung-Sik Lee, Da-Chuan Lu, and Xiaoyang Huang for helpful discussions and/or comments on a draft version of this article. Research
at Perimeter Institute is supported in part by the Government of Canada through the
Department of Innovation, Science and Economic Development and by the Province of
Ontario through the Ministry of Colleges and Universities.

    \begin{appendices}
    \section{Classification of $\mathrm{LU}(1)$ SPTs}
    \label{appendix:classification}
        In this appendix, we will discuss the formal classification of SPT phases with $\mathrm{LU}(1)$ symmetry. Under the hypothesis that SPT phases in $d+1$ spatial dimensions are in one-to-one correspondence with 't Hooft anomalies in $d$ spatial dimensions, this should allow us to describe 't Hooft anomalies for $\mathrm{LU}(1)$ symmetries.
    
    We will make the following general assumption. Let $G$ be a topological group (which includes finite and infinite-dimensional Lie groups as special cases). Then we assume that there exists a so-called ``generalized cohomology theory'' $h^{\bullet}$ such that SPT phases in $d$ spatial dimensions with symmetry $G$ are in one-to-one correspondence with $h^d(BG)$. One can argue for this result based on the assumption that SPT phases are classified by (invertible) topological quantum field theories, e.g. see Ref.~\cite{Freed_1604}; see also Refs.~\cite{Xiong_1701,Gaiotto_1712}.

    Next we invoke the following fact \cite{Atiyah__1983}: $B\mathrm{LU}(1)$ is homotopy equivalent to $LB \mathrm{U}(1)$, i.e.\ the free loop space of $B\mathrm{U}(1)$ (i.e. the space of all loops in the space $\mathrm{BU}(1)$). Homotopy-equivalent spaces have the same generalized cohomology, i.e.\ $h^d(B\mathrm{LU}(1)) \cong h^d(LB\mathrm{U}(1))$.
    Then we observe that there is a natural map $\mathcal{D} : S^1 \times LB\mathrm{U}(1) \to B\mathrm{U}(1)$, under which $(s,\gamma) \mapsto \gamma(s)$, where $s \in S^1$ and $\gamma$ is a loop in $LB\mathrm{U}(1)$, i.e. a continuous function $\gamma : S^1 \to B\mathrm{U}(1)$. Any continuous function $f : X \to Y$ between topological spaces induces a homomorphism in the opposite direction on the generalized cohomology, i.e. a homomoprhism $f^* : h^d(Y) \to h^d(X)$. In particular, $\mathcal{D}$ induces a homomorphism $\mathcal{D}^* : h^d(B\mathrm{U}(1)) \to h^d(S^1 \times B\mathrm{LU}(1))$. Finally, one can show from the axioms of generalized cohomology that $h^d(S^1 \times B\mathrm{LU}(1)) \cong h^d(B\mathrm{LU}(1)) \times h^{d-1}(B\mathrm{LU}(1))$. Projecting onto the second factor, we have constructed a homomorphism from $h^d(B\mathrm{U}(1)) \to h^{d-1}(B \mathrm{LU}(1))$.

    \end{appendices}

    \printbibliography


\end{document}